# A MoS$_2$-based capacitive displacement sensor for DNA sequencing

A. Smolyanitsky,[1*] B. I. Yakobson,[2] T. A. Wassenaar,[3,4] E. Paulechka,[1] K. Kroenlein[1]

[1]Applied Chemicals and Materials Division, National Institute of Standards and Technology,

Boulder, CO 80305, USA

[2]Department of Materials Science and NanoEngineering, Rice University,

Houston, TX 77005, USA

[3]Groningen Biomolecular Sciences and Biotechnology Institute and Zernike Institute for
Advanced Materials, University of Groningen,

9747 AG Groningen, the Netherlands

[4]Bioinformatics, Hanze University of Applied Sciences,

9747 AS Groningen, the Netherlands

[*]To whom correspondence should be addressed: alex.smolyanitsky@nist.gov

**We propose an aqueous functionalized molybdenum disulfide nanoribbon suspended over a solid electrode as the first capacitive displacement sensor aimed at determining the DNA sequence. The detectable sequencing events arise from the combination of Watson-Crick base-pairing, one of nature's most basic lock-and-key binding mechanisms, with the ability of appropriately sized atomically thin membranes to flex substantially in response to sub-nanonewton forces. We employ carefully designed numerical simulations and theoretical estimates to demonstrate excellent (79 % to 86 %) raw target detection accuracy at ~70 million bases per second and electrical measurability of the detected events. In addition, we demonstrate reliable detection of repeated DNA motifs. Finally, we argue that the use of a nanoscale opening (nanopore) is not requisite for the operation of the proposed sensor and present a simplified sensor geometry without the nanopore as part of the sensing element. Our results therefore potentially suggest a realistic, inherently base-specific, high-throughput electronic DNA sequencing device as a cost-effective *de-novo* alternative to the existing methods.**

**Introduction**

For the past two decades, nanotechnology-based DNA sequencing methods have been an area of intense research, first aimed at providing a fast, accurate, and inexpensive alternative to the slow and costly Sanger method[1] and, more recently, to the now-ubiquitous sequencing by synthesis[2] still limited by equipment cost and throughput. Starting with the pioneering work by Kasianowicz *et al.* on utilizing ion current blockage in nanopores for detecting individual nucleotides,[3] a wide variety of nanoscale ionic sequencers have been suggested.[4] Because single-measurement error rates in ion-blocking methods can be as high as 90 %,[4] further research has been focused on developing yet alternative approaches with higher single-measurement accuracy, thus reducing the need for repeated measurements and data post-processing. Such alternatives have ranged from measuring tunneling currents via base-pair hydrogen bonds[5] to using graphene nanopores in ionic sequencers.[6-9] In an intriguing departure from the ion current measurement approach, graphene-based field-effect transistors with nucleotide-specific electronic response were proposed.[10-14]

Although these approaches show promise, thermally induced noise and device scaling issues remain the most significant challenges in the nanopore-based sequencing methods in general,[15] while most of the theoretically described field-effect based devices assume operational temperature near zero kelvin. Aiming for a realistic and naturally nucleotide-specific sequencer not relying on either ionic currents, or field effects, we recently simulated a strain-sensitive graphene nanoribbon (GNR) at room temperature in aqueous environment.[16] As proposed, a single-strand DNA (ssDNA) molecule is translocated via a nanopore in a locally suspended GNR at a given rate. The interior of the nanopore is chemically functionalized with a nucleobase complementary to the target base subject to detection.[16] As target ssDNA bases pass, Watson-Crick base-pairing temporarily deflects the nanoribbon out of plane, in turn causing changes in the GNR conductance via near-uniaxial lattice strain. A single-measurement sequencing accuracy in the vicinity of 90 % without false positives was estimated for the G-C pair at the effective sequencing rate of ~66 million nucleotides per second.

As previously noted,[16] the so-called π-π stacking, effectively resulting in DNA adsorption on pristine graphene, presents a challenge for insertion and translocation of the DNA strand subject to sequencing. Although engineering graphene's hydrophobicity via local non-covalent coating is possible to alleviate the issue of adsorption,[17] replacing graphene with a significantly less hydrophobic atomically thin membrane is a highly attractive option. Molybdenum disulfide ($MoS_2$) is an excellent candidate, because it has been shown to be a non-DNA-adsorbing atomically thin material in the ionic sequencing approach.[18-19]

In this work, we combine Density Functional Theory (DFT) simulations, room temperature molecular dynamics (MD) simulations, and analytical calculations to investigate the operation of a nucleobase-functionalized monolayer $MoS_2$ nanoribbon as a central element in a displacement sensor aimed at selective detection of nucleotides. In contrast with relying on the response of graphene's electronic properties to lattice strain,[16] here we propose a nanoscale flat-plate capacitor, in which one of the plates is selectively deflected out of plane by the passing target nucleotides during DNA translocation. The sequencing readout is then performed as a measurement of the time-varying capacitance. In addition, as shown further, the relatively high bending rigidity of $MoS_2$[20-21] results in significantly reduced flexural fluctuations, compared to graphene, potentially reducing the amount of readout signal noise. At the same time, the flexibility of monolayer $MoS_2$ is shown to be sufficient to allow considerable out-of-plane nanoribbon deformation in response to the forces required to break up a Watson-Crick pair.

Because functionalization of MoS$_2$ with organic molecules has been experimentally demonstrated,[22-23] there exists a realistic possibility of an experimental implementation of the proposed approach.

**System description**

The proposed sequencer aimed at detecting guanine (G) base is sketched in Fig. 1 (a). As shown, the interior of the pore formed in the MoS$_2$ nanoribbon is functionalized by cytosine (C) molecules, which are complementary to G. The metal electrode at the bottom of the proposed sensor forms a flat-plate capacitor with the locally suspended monolayer MoS$_2$ nanoribbon. In such a setup, the modification of capacitance caused by the temporary deflection of the nanoribbon is subject to measurement, as mentioned earlier and depicted in Fig. 1 (b). As shown further, the capacitance variation in response to the ribbon deflections and the resulting electrical signal are measurable using existing integrated circuits without requiring microscopy methods. Following the Watson-Crick base-pairing principle, the "raw" (single-read) DNA sequence can then be obtained using at least two different strategies. In one, the sequence is produced in a single DNA translocation via a stack of four sensors (*e.g.* cytosine-functionalized nanoribbon aimed at detecting guanine and vice versa, *etc.*). Alternatively, the sequence may be constructed from simultaneous scans of identical DNA copies via four sensors, each aimed at a single base type. In principle, the presented displacement sensor is expected to be applicable to all sufficiently flexible, electrically conductive (under appropriately selected bias) membranes, including graphene. Importantly, as discussed further, alternative geometries are also possible in this approach, potentially eliminating the need for the nanopore in the main sensing element.

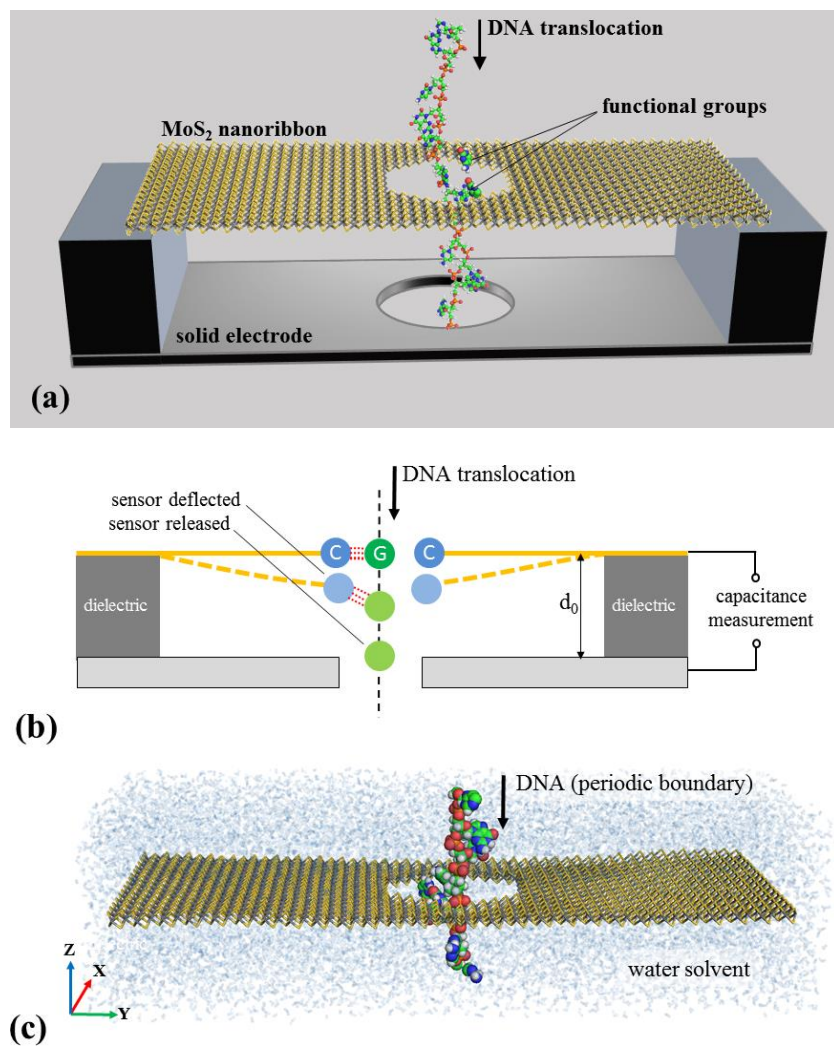

Figure 1. A 3-D sketch of the proposed capacitive displacement sensor (a), principle of operation of the sensor (b), and the complete atomistic system immersed in water, as simulated (c). The red dotted lines in (b) represent hydrogen bonds.

The system subject to MD simulations is shown in Fig. 1 (c). The interior of the pore in the MoS$_2$ nanoribbon is functionalized by two cytosine molecules. Functionalization with a cytosine moiety was achieved via a single covalent S-C bond with the cytosine carbon at position six. The orientation of the functional group relative to MoS$_2$ plane was confirmed by DFT energy minimization, as detailed in the Supplementary Information (SI) and in the Methods section. The DFT simulations were performed on a system consisting of a triangular monolayer MoS$_2$ cluster with a cytosine molecule attached as shown in Fig. S1 of SI. In the MD simulations, the nanoribbon dimensions are $L_x = 4.5$ nm $\times$ $L_y = 15.5$ nm; the nanopore diameter is ~2.5 nm. The ends of the nanoribbon were position-restrained so as to mimic local binding to the supporting substrate (see Fig. 1 (a)). Each simulated ssDNA sample consisted of six bases. In order to reduce the computational cost and enable continuous ssDNA translocation, each DNA strand was made periodic in the Z-direction, as shown in Fig. 1 (c). Prior to production simulations, periodic ssDNA samples were pre-stretched along Z-direction. A total of six potassium ions were added to the solvent to counteract the negative net charge of the six-base DNA samples. Similarly to previous work,[16] weak in-plane harmonic position restraints with a constant of $200.0 \frac{kJ}{mol\ nm^2}$ were applied to the six CH$_2$-bound oxygens of the phosphate moieties, mimicking the effect of an insertion aperture, which maintains the DNA position reasonably close to the center of the nanopore, while allowing rotation around Z-axis.

**Methods**

The DFT simulations aimed at determining the stability of the functional group (cytosine) and its orientation relative to the MoS$_2$ plane were performed using the CP2K package.[24] Perdew, Burke and Ernzerhof (PBE) exchange functional,[25] Gaussian plane-wave pseudopotentials,[26-27] and the DZVP basis set[28] were used. In addition, D3 non-local correction[29] was applied. All MD simulations were performed using GROMACS 5.1.2 package.[30] The MD models of the DNA and functionalized MoS$_2$ were based on the AMBER94 forcefield.[31] The intramolecular interactions in MoS$_2$ were set according to previous work[32] and further refined to reproduce the basic mechanical material properties in a reasonable manner (for further details, see section S2 of SI). The charges of sulfur and molybdenum atoms were set according to quantum-mechanical calculations.[33] The system was immersed in a rectangular container filled with explicit water molecules, using the TIP4P model.[34-35] Prior to the production MD simulations, all systems underwent *NPT* relaxation at $T = 300$ K and $p = 0.1$ MPa. The production simulations of the DNA translocation via nanopores were performed in an *NVT* ensemble at $T = 300$ K, maintained by a velocity-rescaling thermostat[36] with a time constant of 0.1 ps.

**Results and discussion**

The results of simulated ssDNA translocation via a functionalized MoS$_2$ nanoribbon are discussed next. In order to assess selective hydrogen bond formation between the functional groups and the target (G) nucleotides, as well as the resulting nanoribbon deflections, a sample sequence *TGAAGC* was set up as shown in Fig. 1 (c) and translocated for 300 ns at an average prescribed rate of 5 cm/s in the negative Z-direction. At the given rate and simulated time, the DNA travels 15 nm along the prescribed direction. Therefore, given a periodic boundary in the Z-direction and the fact that the pre-stretched six-base DNA sample length was approximately 4.4 nm along the Z-axis, the sample sequence is expected to traverse the pore 15 nm / 4.4 nm ≈ 3.4 times. Therefore, the complete test sequence, as seen by the functional groups in the nanopore, was close to *TGAAGC*/*TGAAGC*/*TGAAGC*/*TG* (underlined base inside the pore at the start of the simulation) with a total of seven guanine passages expected. The nanoribbon deflection data (maximum deflection at the nanoribbon center and average deflection $\langle h \rangle = \frac{1}{N_{Mo}} \sum_{N_{Mo}} z_i$ calculated from a total of $N_{Mo}$ molybdenum atoms), together with the number of hydrogen bonds as functions of simulated time, are shown in Fig. 2 (a). From the hydrogen bond formation data, seven binding events indeed occur, as enumerated in Fig. 2 (a). With the exception of $G_2$, for which the duration of binding is the shortest, all hydrogen bond formation events are accompanied by deflection events beyond the provided thresholds. At the same time, no false-positive deflections beyond thresholds occur, which suggests an overall raw detection error in the vicinity of one out of seven, or 14 %. One notes that the deflections are significantly lower than those reported for a graphene nanoribbon of similar dimensions described earlier.[16] This result owes to the significantly higher bending rigidity of $MoS_2$, compared to graphene.[20-21] The vertical force causing selective deflections can be evaluated directly from Fig. 2 (b), where the DNA external pulling force is plotted as a function of simulated time. At the peaks corresponding to the deflection maxima, the critical force required to break up the resulting G-C pairs is obtained. From averaging over six "useful" deflection events, the force peak magnitude is ≈ 60 pN, in good agreement with previous results[16] and experimental data.[37-38]

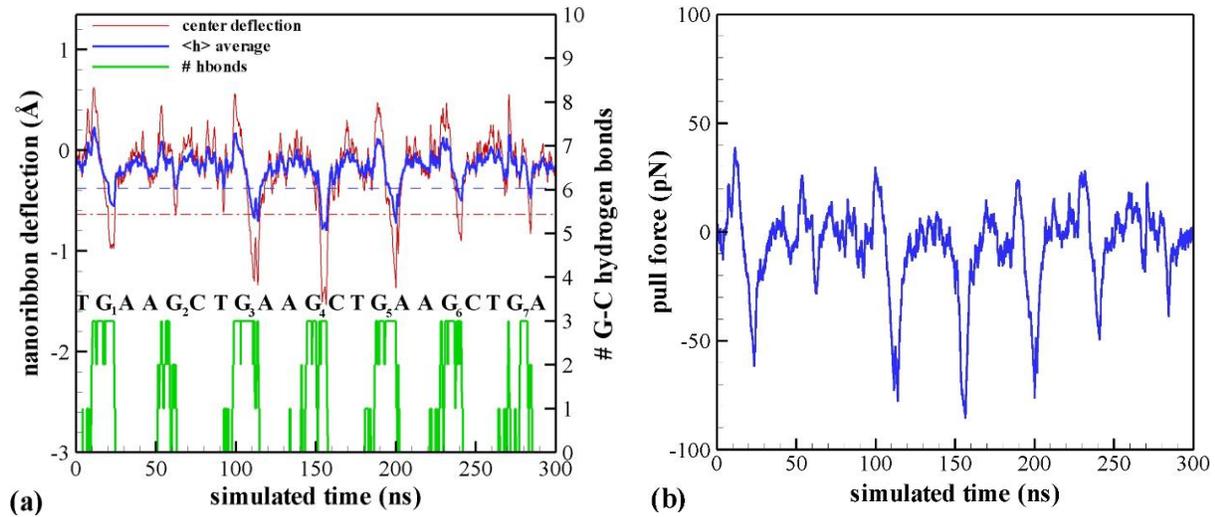

Figure 2. Maximum (center) and average deflection $\langle h \rangle$ of the $MoS_2$ nanoribbon, along with the number of G-C hydrogen bonds (a), and ssDNA pulling force (b) as functions of simulated time. The DNA translocation rate was 5 cm/s in the negative Z-direction. A low-pass filter with 800 MHz cutoff was applied to the raw deflection and pulling force data. The red and blue horizontal lines in (a) represent basic thresholds to guide visual inspection of the useful deflection events for the center deflection and $\langle h \rangle$, respectively.

Experimental detectability of the deflection events is critical for the DNA sequencing application. In the capacitive sensor scheme proposed here, the relative change in capacitance is straightforward to estimate as $\frac{\Delta C}{C_0} \approx -\frac{\langle h \rangle}{d_0}$ (see SI for the derivation), reasonably assuming $\langle h \rangle \ll d_0$, where $d_0$ is the plate separation, as defined in Fig. 1 (b). The value of $\langle h \rangle$ averaged over the six deflection events in Fig. 2 (a) is $\langle h \rangle_{ave} \approx 0.6$ Å, and thus with $d_0 = 1.0$ nm, $\frac{\Delta C}{C_0} \approx 6$%. The baseline capacitance $C_0$ (see Eq. (S1) in SI) for even the small nanoribbon in this work yields ~ 53.1 aF, experimentally measurable on-chip in an AC measurement.[39-40] Alternatively, simulated polarization of the DNA molecule itself in response to a rapidly alternating high-amplitude electric field was previously proposed for determining nucleotide species in an AC measurement.[41] However, here the nature of time dependence of the capacitance resulting from membrane deflections shown in Fig. 2 (a) allows detection of individual deflection events using an integrated DC circuit. The detailed discussion of the proposed measurement strategy is as follows. Consider the equivalent circuit representing the sensor, shown in Fig. 3 (a). Due to the possible presence of electrolyte ions in the aqueous system containing DNA, an ionic conductor is connected in parallel with the ideal capacitor formed between the MoS$_2$ membrane and the solid electrode sketched in Fig. 1 (a). An appropriately selected constant voltage $V_0$ is applied across the sensor and the total current through the circuit is the effective measured signal, which is fed to the amplifier stage as a voltage drop across a small resistive load $R_0$, as shown in Fig 3 (a). An additional noise voltage contribution $\delta V \ll V_0$ is also present in the system, as discussed further. Only first-order perturbative effects are considered here.

The total current in the circuit is $I_{tot}(t) = [i_d(t) + i'_d(t)] + i_i(t)$, where $i_d(t) = V_0 \frac{dC(t)}{dt}$ (with $C(t) \approx C_0 \left(1 - \frac{\langle h \rangle(t)}{d_0}\right)$, as estimated in section S3 of SI) is the displacement current associated with membrane deflections, $i'_d(t) = C_0 \frac{d\delta V(t)}{dt}$ is the displacement current noise from voltage perturbations $\delta V(t)$ contributed by the solvent, dissolved ions, as well as the ssDNA, and $i_i(t)$ is the ionic leakage current, also subject to perturbation due to varying electric field between the capacitor plates. Here, we assume that most of the "useful" plate charge perturbation is contributed by the change in the capacitor geometry due to membrane deflections, while the density of mobile charge carriers in the semiconducting MoS$_2$ ribbon remains constant.

Given the definitions above, a data excerpt from the simulation that yielded the results in Fig. 2 was used directly to reveal detailed time dependence of the electrical response to membrane deflections. In particular, $C(t)$ and $i_d(t)$ are plotted in Fig. 3 (b) for $V_0$ = 150 mV (see section S5 of the SI). As expected, $i_d(t)$ oscillates around zero overall and produces pairs of transient peaks in excess of 50 pA when the membrane deflects and slips back. In absence of other contributions, these current spikes represent the primary signal subject to detection. As estimated, the 50 pA transient current amplitude at the given timescale is within the existing measurement capabilities [42-43] even for the small membrane considered here.

The noise arising from fast fluctuations of the solvent and the dissolved ions is expected to be in the frequency range far beyond that of the "useful" signal. However, the electrostatic bias noise due to the motion of the ssDNA sample, including its translocation and any spurious movements, occurs within the timescale of interest. Conveniently, the noise current $i'_d(t) = C_0 \frac{d\delta V(t)}{dt}$ can be estimated directly from the simulated electrostatics. We note that $\delta V(t)$ can be obtained from the time-dependent solution of the Poisson's equation in the region occupied by the MoS$_2$ membrane, as contributed by the DNA atomic charges. As shown in Fig. 3 (c), $\delta V(t)$ indeed varies relatively slowly during DNA translocation and the resulting displacement current noise $i'_d(t)$ amplitude is only 10 % to 15 % of the $i_d(t)$ peaks in Fig. 3 (b). Importantly, this noise contribution is expected to further decrease with increasing membrane size due to the $\sim 1/r$ dependence of the electrostatic potential perturbations contributed by a near-linear strand of DNA perpendicular to the membrane.

Finally, the ionic leakage current $i_i(t)$ and the total current $I_{tot}(t) = [i_d(t) + i'_d(t)] + i_i(t)$ through the circuit are estimated. The ionic current between the capacitor plates of length $L$ and width $w$ (assuming the "worst-case" scenario, in which each ion transfers charge to the membrane) is estimated for dissolved KCl as $i_i(t) = \frac{nwLqV_0(\mu_K+\mu_{Cl})}{d_0}\left(1 - \frac{\langle h \rangle(t)}{d_0}\right)$, where $n$, $\mu_K$, and $\mu_{Cl}$ are the electrolyte concentration and the ionic mobilities, respectively (see section S4 of SI for details). A 5 mM KCl concentration is assumed. As shown in Fig. 3 (d), the ionic contribution results in a significant overall current baseline, subject to transient fluctuation via $\frac{\langle h \rangle(t)}{d_0}$. Importantly, however, deflection-induced *variation* of the total current $I_{tot}(t)$ remains dominated by the displacement current $i_d(t)$ for the selected salt concentration. It is then clear that further increasing electrolyte concentration would eventually mask the capacitive effect entirely. The presence of electrolyte ions in the system suggests a potentially more serious challenge for this system, as well as any sensor concept, which relies on the mechanical and/or electronic properties of the atomically thin membranes. Although little is known about ion adsorption on MoS$_2$ in aqueous environment, electrochemical material deposition on the membrane surface may occur, potentially leading to significant changes of the properties of the resulting composite during sensor operation. Therefore, de-ionization of the DNA samples,[44] membrane passivation, and/or providing an alternative conductive path for the mobile electrolyte ions via additional fields may be considered to address this challenge (also see section S6 of SI).

Because bending properties of the ribbon material are known, along with a reasonable estimate of the pulling force arising from splitting base-pairs, both $\langle h \rangle$ ($\propto L^3/w$) and $C_0 \propto Lw/d_0$ are subject to refined design in terms of the ribbon dimensions. The value of $d_0$ (and thus the bias voltage $V_0$) should then also be optimizable for larger nanoribbons to achieve optimal signal contributions, while remaining within the reach of device fabrication capability (also see sections S4-S6 of SI).

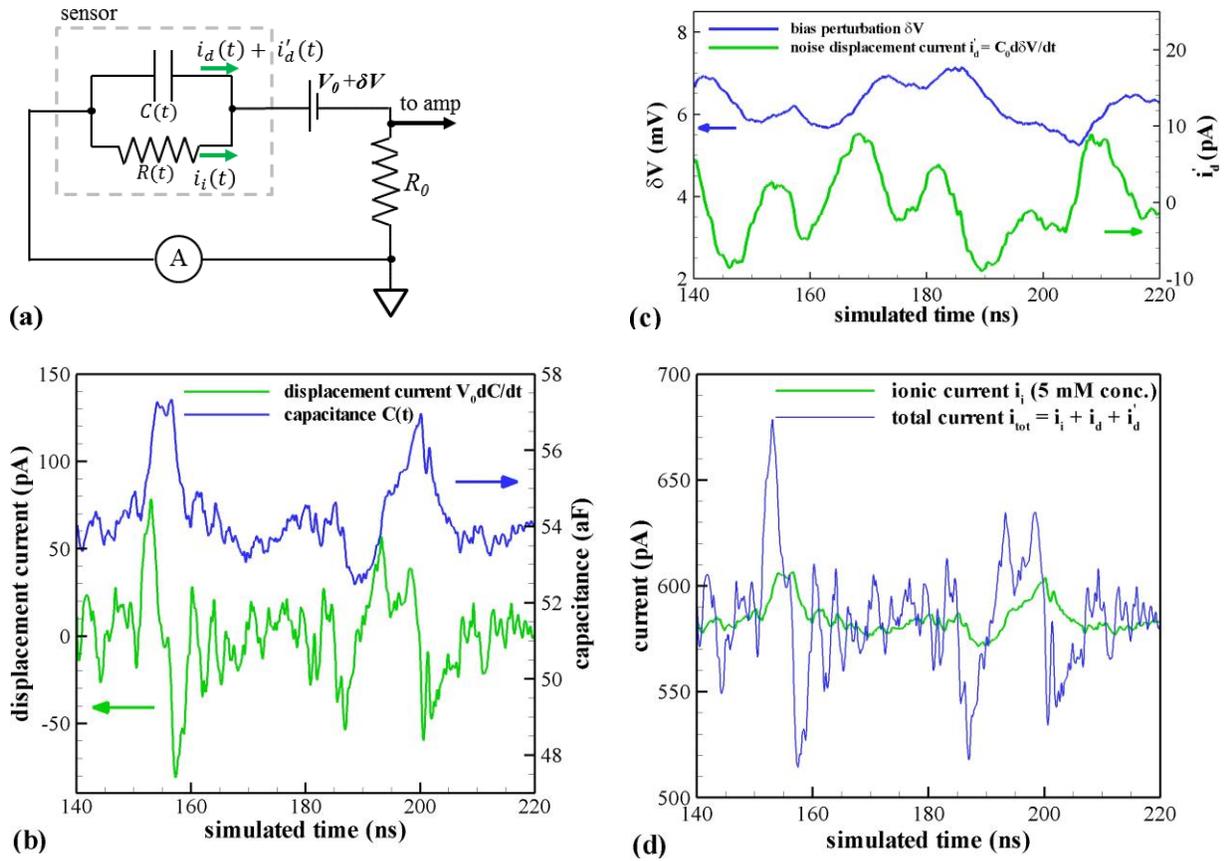

Figure 3. Simplified equivalent circuit of the sensor (a), displacement current and capacitance as functions of simulated time (b), bias perturbation contributed by the translocating DNA sample (c), ionic current contribution and the total current in the circuit (d).

The data presented in Fig. 2 corresponded to a DNA sequence …*TGAAGC*…, in which target guanines were separated by two non-target bases. Given that the proposed detection mechanism relies on hydrogen bond formation and subsequent deflections of the nanoribbon, a realistic motif consisting of repeated target nucleotides may present a sequencing challenge. This challenge is two-fold, including "skipping" detection of the targets due to their close spacing (especially when the expected maximum deflections of a given nanoribbon are comparable to the base spacing), as well as guanine-guanine interactions within an ssDNA chain, which may cause "interference" during interactions with the functional groups at the pore interior. The latter can arise from hydrogen bonding between a hydrogen atom of the amino group and the carboxylic oxygen of the neighboring guanine moieties.

In order to investigate detection of a repeated target sequence and also to provide a comparison with the results obtained for a sequence containing no target bases, additional translocation simulations were set up as described above and run for 200 ns. The results obtained for the test sequences …$GGGGGG$… (all-target) and …$AACCTT$… (non-target) are shown in Fig. 4. For the all-target sequence, 11 distinct deflection events (with an average of $\langle h \rangle_{ave} \approx$ 0.37 Å) are observed, while only thermal fluctuations are observed for the non-target case. The reduction of the average deflection magnitude compared to the results in Fig. 2 (a) is likely attributable to the "interference" effects mentioned above. Irregularities in event periodicity, as well as clearly missed events (*e.g.* between 150 ns and 170 ns) are also present. In 200 ns, a total of 14 complete target base passages are expected and, given that 11 deflection events are observed, the raw detection accuracy, as calculated from the presented data, is 11/14 ≈ 79 %. In order to resolve the presence of a repeated sequence better, we calculated the Fourier spectra of the time-dependent deflection data, as shown in the inset of Fig. 4 (a). In contrast with the spectral distribution obtained for the non-target sequence, an outstanding $f_0 = 72$ MHz peak is observed for the all-target case, corresponding to a base spacing of $v_{scan} \times 1/f_0 = 6.94$ Å. Given the ≈ 4.14 nm length of the periodic pre-stretched all-target sample consisting of six bases along the Z-axis, the event periodicity from a purely geometric standpoint is 4.14 nm / 6 = 6.90 Å, in excellent agreement with the periodicity obtained from the spectrum. Therefore, given that the translocation rate is known, a continuous calculation of the spectral properties of the deflection data (performed within an appropriately selected time "window") can serve as an effective repeated sequence detection measure.

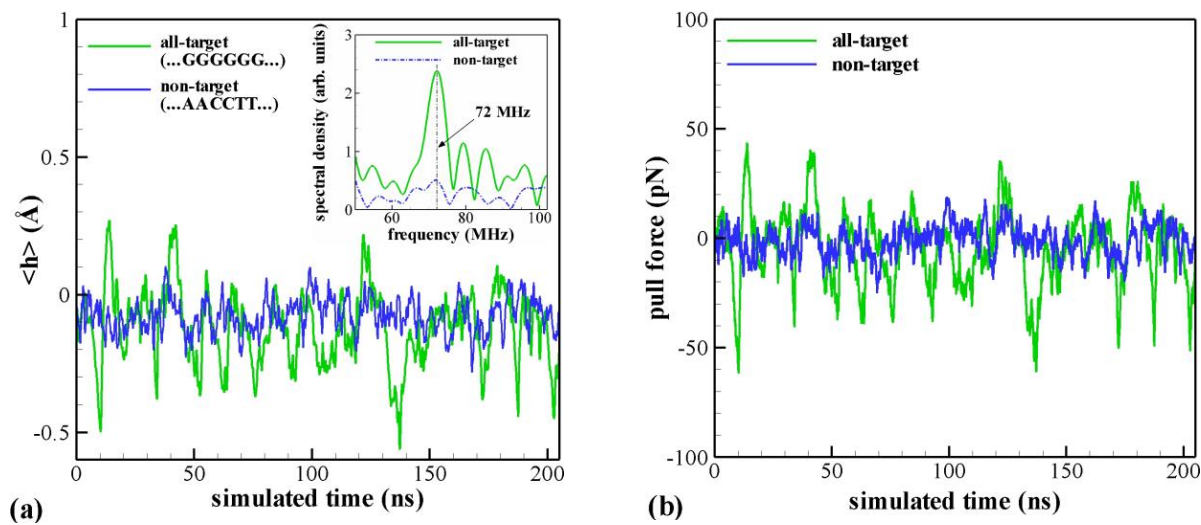

Figure 4. Average nanoribbon deflections $\langle h \rangle$ (a) and ssDNA pulling forces (b) as functions of simulated time for the repeated all-target sequence and the non-target sequence. The inset in (a) shows frequency spectra obtained from the presented time-domain data. The DNA translocation rate was 5 cm/s in the negative Z-direction. A low-pass filter with 800 MHz cutoff was applied to the raw deflection data.

The results presented in Figs. 2 and 4 were obtained for the DNA translocation rate of 5 cm/s (corresponding to the read rate 14 ns/base or ~70 million bases per second), as dictated by the computational load associated with performing long MD simulations of a relatively large system with explicit solvent. As shown in the discussion accompanying Fig. 4, the useful signal frequency range associated with the 5 cm/s translocation rate is well within the capacity of the currently available measurement equipment. At the same time, some of the fastest experimental readouts for the ionic current based methods correspond to 1-3 μs/base,[45] owing in part to the limitations of measuring fast-changing ionic currents. Although the approach proposed here does not rely primarily on ionic currents (and thus not subject to the limitations associated with their measurement) and MD simulations of DNA translocation at microseconds per base are beyond our current computational capability, we performed an additional 1.2 μs long ssDNA translocation simulation at 1 cm/s, corresponding to 70 ns/base or 14 million bases per second. For the DNA sequence identical to that in Fig. 2, the results are presented in Fig. 5. With the exception of the short binding event at ~0.75 μs, distinct nanoribbon deflections accompany all of the target binding events, similar to the results in Fig. 2. Therefore, translocation rate reduction by a factor of five does not appear to degrade target detection rate.

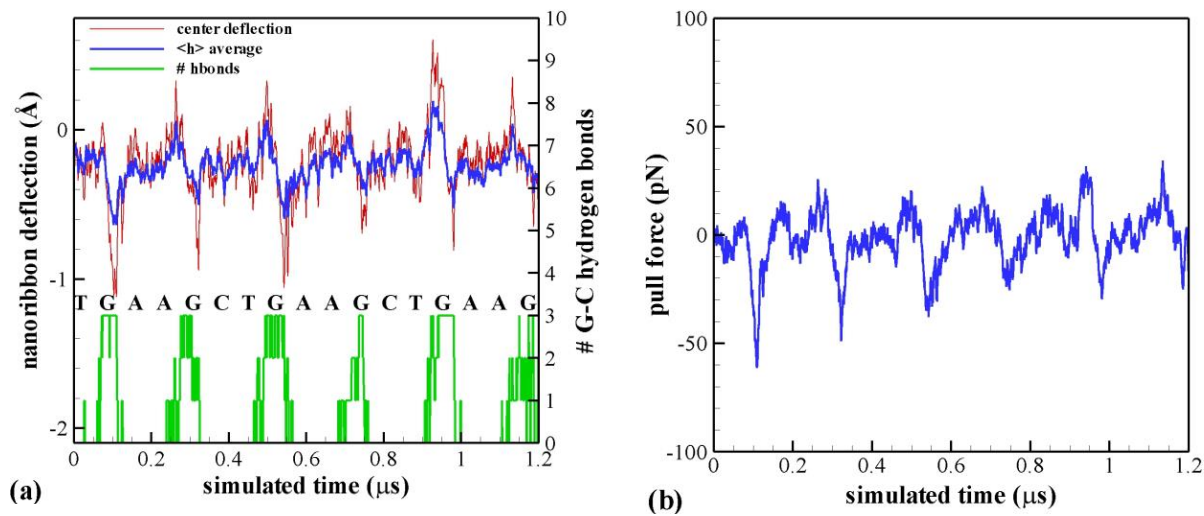

Figure 5. Maximum (center) and average deflection ⟨h⟩ of the MoS$_2$ nanoribbon, along with the number of G-C hydrogen bonds (a), and ssDNA pulling force (b) as functions of simulated time. The DNA translocation rate was 1 cm/s in the negative Z-direction. A low-pass filter with 200 MHz cutoff was applied to the raw deflection data.

An ever-present challenge associated with all nanopore-based sequencing methods is precise insertion of the DNA strand into the narrow pore, followed by DNA translocation with minimal amount of spurious motions. The latter can be especially important for high translocation rates, highly desirable for the proposed ultra-high-speed sequencer. A unique feature of the sequencing approach described both here and in the previous work,[16] however, is that the presence of a nanopore in the sensor membrane itself is not required. A simpler and possibly more realistic alternative in terms of fabrication, functionalization, and setup is presented in Fig. 6 (a), where the ssDNA sample is shown to be translocated perpendicularly to the functionalized edge of the locally suspended membrane, omitting the nanopore entirely. Such a geometry still requires a solid aperture for proper positioning of the DNA sample relative to the sensor, but eliminates the need for carving a nanopore in an atomically thin membrane, as well as the need for molecular functionalization in a highly confined region. In this configuration, a twisting deformation would be caused in addition to bending and stretching, possibly modifying the useful signal estimates for the readout scheme previously proposed for graphene.[16] However, for the capacitive readout mechanism proposed in this work, the relative change in capacitance due to deflection is $\frac{\Delta C}{C_0} \approx -\frac{\langle h \rangle}{d_0}$, which is not sensitive to possible additional twisting, as long as $\langle h \rangle$ is nonzero, expected for a suspended nanoribbon. The distribution of the out-of-plane atomic positions throughout the membrane is shown in Fig. 6 (b), as obtained for a $F_z = 75$ pN out-of-plane force applied at the edge. Although some degree of twisting is observed, the membrane is deflected throughout, with $\langle h \rangle \approx 0.6$ Å.

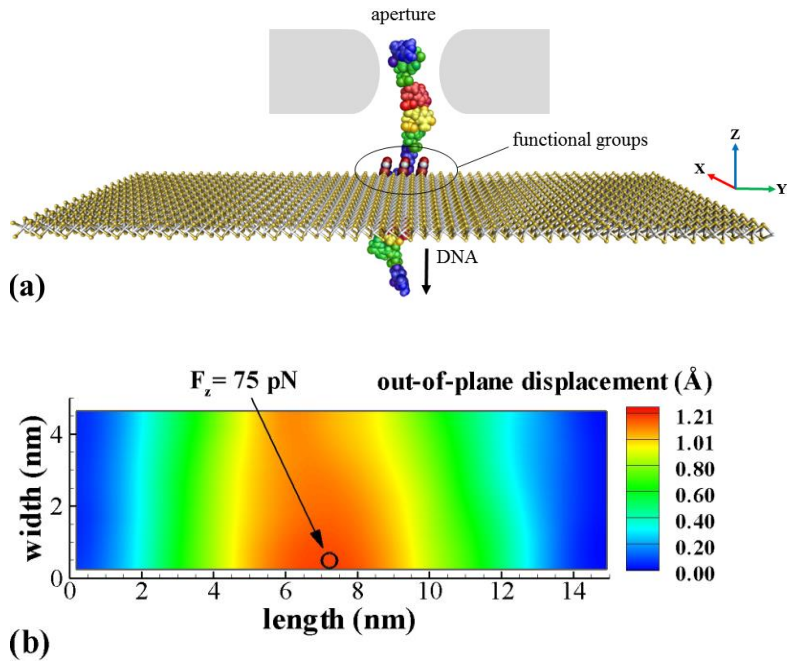

Figure 6. Edge sensor configuration without the nanopore (a) and the distribution of the out-of-plane atomic positions throughout the membrane (b), obtained for a constant $F_z = 75$ pN force applied as shown.

**Conclusions**

We have proposed a nucleobase-functionalized $MoS_2$ nanoribbon suspended over a solid metal electrode as a capacitive displacement sensor for ultra-fast and accurate DNA sequencing at room temperature. The proposed sensing mechanism combines Watson-Crick base-pairing with the ability of nanoscale atomically thin membranes to flex in response to sub-nanonewton forces. Unlike graphene, $MoS_2$ is a non-DNA adsorbing material, which effectively resolves adsorption-related issues outlined earlier.[16] A raw (single-read) sequencing accuracy in the vicinity of 79-86 % is demonstrated for the translocation rates ranging from 14 to 70 million bases per second. Even for the relatively small nanoribbons simulated here, electronic measurement of the target base detection events is estimated to be electrically measurable. Further device size optimization is possible in terms of fabrication and improved measurability of the deflection-induced sequencing events. In addition, we confirm detection of repeated target base sequences and show that Fourier analysis of the deflection data is a useful repeated motif detection measure. Finally, we argue that the presence of a nanopore in the membrane may not be required for the sequencing approaches presented both here and in our previous work,[16] and present an alternative geometry, in which the DNA is translocated perpendicularly to the edge of a locally suspended nanoribbon without a pore. The proposed sensing approach therefore holds promise for a realistic, accurate, and ultra-fast DNA sequencing technology.

**Acknowledgements**

The authors are grateful to A. Balijepalli, J. Suehle, A. Davydov, Y. Shevchenko, and N. Dyubankova for illuminating discussions. NIST authors gratefully acknowledge support from the Materials Genome Initiative. B.I.Y. at Rice University was supported by the Office of Naval Research grant N00014-15-1-2372 and by the Robert Welch Foundation (C-1590). T.A.W. acknowledges support by S.J. Marrink and the generous computational support by the Donald Smit Center at the University of Groningen. A.S. acknowledges computational support by the NIST Center for Theoretical and Computational Materials Science and personally thanks A.C.E. Reid. The authors also thank A.F. Kazakov for the help with visualization.

This work is a contribution of the National Institute of Standards and Technology, an agency of the US government. Not subject to copyright in the USA. Trade names are provided only to specify procedures adequately and do not imply endorsement by the National Institute of Standards and Technology. Similar products by other manufacturers may be found to work as well or better.

# References

1. Sanger, F.; Nicklen, S.; Coulson, A. R., DNA sequencing with chain-terminating inhibitors. *PNAS* **1977,** *74* (12), 5463-5467.
2. Bentley, D. R.; Balasubramanian, S.; Swerdlow, H. P.; Smith, G. P.; Milton, J.; Brown, C. G.; Hall, K. P.; Evers, D. J.; Barnes, C. L.; Bignell, H. R.; Boutell, J. M.; Bryant, J.; Carter, R. J.; Keira Cheetham, R.; Cox, A. J.; Ellis, D. J.; Flatbush, M. R.; Gormley, N. A.; Humphray, S. J.; Irving, L. J.; Karbelashvili, M. S.; Kirk, S. M.; Li, H.; Liu, X.; Maisinger, K. S.; Murray, L. J.; Obradovic, B.; Ost, T.; Parkinson, M. L.; Pratt, M. R.; Rasolonjatovo, I. M. J.; Reed, M. T.; Rigatti, R.; Rodighiero, C.; Ross, M. T.; Sabot, A.; Sankar, S. V.; Scally, A.; Schroth, G. P.; Smith, M. E.; Smith, V. P.; Spiridou, A.; Torrance, P. E.; Tzonev, S. S.; Vermaas, E. H.; Walter, K.; Wu, X.; Zhang, L.; Alam, M. D.; Anastasi, C.; Aniebo, I. C.; Bailey, D. M. D.; Bancarz, I. R.; Banerjee, S.; Barbour, S. G.; Baybayan, P. A.; Benoit, V. A.; Benson, K. F.; Bevis, C.; Black, P. J.; Boodhun, A.; Brennan, J. S.; Bridgham, J. A.; Brown, R. C.; Brown, A. A.; Buermann, D. H.; Bundu, A. A.; Burrows, J. C.; Carter, N. P.; Castillo, N.; Chiara E. Catenazzi, M.; Chang, S.; Neil Cooley, R.; Crake, N. R.; Dada, O. O.; Diakoumakos, K. D.; Dominguez-Fernandez, B.; Earnshaw, D. J.; Egbujor, U. C.; Elmore, D. W.; Etchin, S. S.; Ewan, M. R.; Fedurco, M.; Fraser, L. J.; Fuentes Fajardo, K. V.; Scott Furey, W.; George, D.; Gietzen, K. J.; Goddard, C. P.; Golda, G. S.; Granieri, P. A.; Green, D. E.; Gustafson, D. L.; Hansen, N. F.; Harnish, K.; Haudenschild, C. D.; Heyer, N. I.; Hims, M. M.; Ho, J. T.; Horgan, A. M.; Hoschler, K.; Hurwitz, S.; Ivanov, D. V.; Johnson, M. Q.; James, T.; Huw Jones, T. A.; Kang, G.-D.; Kerelska, T. H.; Kersey, A. D.; Khrebtukova, I.; Kindwall, A. P.; Kingsbury, Z.; Kokko-Gonzales, P. I.; Kumar, A.; Laurent, M. A.; Lawley, C. T.; Lee, S. E.; Lee, X.; Liao, A. K.; Loch, J. A.; Lok, M.; Luo, S.; Mammen, R. M.; Martin, J. W.; McCauley, P. G.; McNitt, P.; Mehta, P.; Moon, K. W.; Mullens, J. W.; Newington, T.; Ning, Z.; Ling Ng, B.; Novo, S. M.; O'Neill, M. J.; Osborne, M. A.; Osnowski, A.; Ostadan, O.; Paraschos, L. L.; Pickering, L.; Pike, A. C.; Pike, A. C.; Chris Pinkard, D.; Pliskin, D. P.; Podhasky, J.; Quijano, V. J.; Raczy, C.; Rae, V. H.; Rawlings, S. R.; Chiva Rodriguez, A.; Roe, P. M.; Rogers, J.; Rogert Bacigalupo, M. C.; Romanov, N.; Romieu, A.; Roth, R. K.; Rourke, N. J.; Ruediger, S. T.; Rusman, E.; Sanches-Kuiper, R. M.; Schenker, M. R.; Seoane, J. M.; Shaw, R. J.; Shiver, M. K.; Short, S. W.; Sizto, N. L.; Sluis, J. P.; Smith, M. A.; Ernest Sohna Sohna, J.; Spence, E. J.; Stevens, K.; Sutton, N.; Szajkowski, L.; Tregidgo, C. L.; Turcatti, G.; vandeVondele, S.; Verhovsky, Y.; Virk, S. M.; Wakelin, S.; Walcott, G. C.; Wang, J.; Worsley, G. J.; Yan, J.; Yau, L.; Zuerlein, M.; Rogers, J.; Mullikin, J. C.; Hurles, M. E.; McCooke, N. J.; West, J. S.; Oaks, F. L.; Lundberg, P. L.; Klenerman, D.; Durbin, R.; Smith, A. J., Accurate whole human genome sequencing using reversible terminator chemistry. *Nature* **2008,** *456* (7218), 53-59.
3. Kasianowicz, J. J.; Brandin, E.; Branton, D.; Deamer, D. W., Characterization of individual polynucleotide molecules using a membrane channel. *Proceedings of the National Academy of Sciences* **1996,** *93* (24), 13770-13773.
4. Zwolak, M.; Ventra, M. D., Colloquium: Physical approaches to DNA sequencing and detection. *Reviews of Modern Physics* **2008,** *80* (1), 141-165.
5. He, J.; Lin, L.; Zhang, P.; Lindsay, S., Identification of DNA Basepairing via Tunnel-Current Decay. *Nano Letters* **2007,** *7* (12), 3854-3858.
6. Garaj, S.; Hubbard, W.; Reina, A.; Kong, J.; Branton, D.; Golovchenko, J. A., Graphene as a subnanometre trans-electrode membrane. *Nature* **2010,** *467* (7312), 190-193.
7. Wells, D. B.; Belkin, M.; Comer, J.; Aksimentiev, A., Assessing Graphene Nanopores for Sequencing DNA. *Nano Letters* **2012,** *12* (8), 4117-4123.
8. Schneider, G. F.; Kowalczyk, S. W.; Calado, V. E.; Pandraud, G.; Zandbergen, H. W.; Vandersypen, L. M. K.; Dekker, C., DNA Translocation through Graphene Nanopores. *Nano Letters* **2010,** *10* (8), 3163-3167.
9. Merchant, C. A.; Healy, K.; Wanunu, M.; Ray, V.; Peterman, N.; Bartel, J.; Fischbein, M. D.; Venta, K.; Luo, Z.; Johnson, A. T. C.; Drndić, M., DNA Translocation through Graphene Nanopores. *Nano Letters* **2010,** *10* (8), 2915-2921.

10. Min, S. K.; Kim, W. Y.; Cho, Y.; Kim, K. S., Fast DNA sequencing with a graphene-based nanochannel device. *Nat Nano* **2011,** *6* (3), 162-165.
11. Dontschuk, N.; Stacey, A.; Tadich, A.; Rietwyk, K. J.; Schenk, A.; Edmonds, M. T.; Shimoni, O.; Pakes, C. I.; Prawer, S.; Cervenka, J., A graphene field-effect transistor as a molecule-specific probe of DNA nucleobases. *Nat Commun* **2015,** *6*.
12. Nelson, T.; Zhang, B.; Prezhdo, O. V., Detection of Nucleic Acids with Graphene Nanopores: Ab Initio Characterization of a Novel Sequencing Device. *Nano Letters* **2010,** *10* (9), 3237–3242.
13. Postma, H. W. C., Rapid Sequencing of Individual DNA Molecules in Graphene Nanogaps. *Nano Letters* **2010,** *10* (2), 420-425.
14. F. Traversi; C. Raillon; S.M. Benameur; K. Liu; S. Khlybov; M. Tosun; D. Krasnozhon; A. Kis; Radenovic, A., Detecting the translocation of DNA through a nanopore using graphene nanoribbons. *Nat Nano* **2013,** *8* (12), 939-945.
15. Branton, D.; Deamer, D. W.; Marziali, A.; Bayley, H.; Benner, S. A.; Butler, T.; Di Ventra, M.; Garaj, S.; Hibbs, A.; Huang, X.; Jovanovich, S. B.; Krstic, P. S.; Lindsay, S.; Ling, X. S.; Mastrangelo, C. H.; Meller, A.; Oliver, J. S.; Pershin, Y. V.; Ramsey, J. M.; Riehn, R.; Soni, G. V.; Tabard-Cossa, V.; Wanunu, M.; Wiggin, M.; Schloss, J. A., The potential and challenges of nanopore sequencing. *Nat Biotech* **2008,** *26* (10), 1146-1153.
16. Paulechka, E.; Wassenaar, T. A.; Kroenlein, K.; Kazakov, A.; Smolyanitsky, A., Nucleobase-functionalized graphene nanoribbons for accurate high-speed DNA sequencing. *Nanoscale* **2016,** *8* (4), 1861-1867.
17. Schneider, G. F.; Xu, Q.; Hage, S.; Luik, S.; Spoor, J. N. H.; Malladi, S.; Zandbergen, H.; Dekker, C., Tailoring the hydrophobicity of graphene for its use as nanopores for DNA translocation. *Nat Commun* **2013,** *4*.
18. Liu, K.; Feng, J.; Kis, A.; Radenovic, A., Atomically Thin Molybdenum Disulfide Nanopores with High Sensitivity for DNA Translocation. *ACS Nano* **2014,** *8* (3), 2504-2511.
19. Farimani, A. B.; Min, K.; Aluru, N. R., DNA Base Detection Using a Single-Layer MoS2. *ACS Nano* **2014,** *8* (8), 7914-7922.
20. Bertolazzi, S.; Brivio, J.; Kis, A., Stretching and Breaking of Ultrathin MoS2. *ACS Nano* **2011,** *5* (12), 9703-9709.
21. Jin-Wu, J.; Zenan, Q.; Harold, S. P.; Timon, R., Elastic bending modulus of single-layer molybdenum disulfide ($MoS_2$): finite thickness effect. *Nanotechnology* **2013,** *24* (43), 435705.
22. Knirsch, K. C.; Berner, N. C.; Nerl, H. C.; Cucinotta, C. S.; Gholamvand, Z.; McEvoy, N.; Wang, Z.; Abramovic, I.; Vecera, P.; Halik, M.; Sanvito, S.; Duesberg, G. S.; Nicolosi, V.; Hauke, F.; Hirsch, A.; Coleman, J. N.; Backes, C., Basal-Plane Functionalization of Chemically Exfoliated Molybdenum Disulfide by Diazonium Salts. *ACS Nano* **2015,** *9* (6), 6018-6030.
23. Tan, C.; Zhang, H., Two-dimensional transition metal dichalcogenide nanosheet-based composites. *Chemical Society Reviews* **2015,** *44* (9), 2713-2731.
24. Hutter, J.; Iannuzzi, M.; Schiffmann, F.; VandeVondele, J., cp2k: atomistic simulations of condensed matter systems. *Wiley Interdisciplinary Reviews: Computational Molecular Science* **2014,** *4* (1), 15-25.
25. Perdew, J. P.; Burke, K.; Ernzerhof, M., Generalized Gradient Approximation Made Simple. *Physical Review Letters* **1996,** *77* (18), 3865-3868.
26. Goedecker, S.; Teter, M.; Hutter, J., Separable dual-space Gaussian pseudopotentials. *Physical Review B* **1996,** *54* (3), 1703-1710.
27. Hartwigsen, C.; Goedecker, S.; Hutter, J., Relativistic separable dual-space Gaussian pseudopotentials from H to Rn. *Physical Review B* **1998,** *58* (7), 3641-3662.
28. VandeVondele, J.; Hutter, J., Gaussian basis sets for accurate calculations on molecular systems in gas and condensed phases. *The Journal of Chemical Physics* **2007,** *127* (11), 114105.
29. Grimme, S.; Antony, J.; Ehrlich, S.; Krieg, H., A consistent and accurate ab initio parametrization of density functional dispersion correction (DFT-D) for the 94 elements H-Pu. *The Journal of Chemical Physics* **2010,** *132* (15), 154104.

30. Van Der Spoel, D.; Lindahl, E.; Hess, B.; Groenhof, G.; Mark, A. E.; Berendsen, H. J. C., GROMACS: Fast, flexible, and free. *Journal of Computational Chemistry* **2005,** *26* (16), 1701-1718.
31. Cornell, W. D.; Cieplak, P.; Bayly, C. I.; Gould, I. R.; Merz, K. M.; Ferguson, D. M.; Spellmeyer, D. C.; Fox, T.; Caldwell, J. W.; Kollman, P. A., A Second Generation Force Field for the Simulation of Proteins, Nucleic Acids, and Organic Molecules. *Journal of the American Chemical Society* **1995,** *117* (19), 5179-5197.
32. Wakabayashi, N.; Smith, H. G.; Nicklow, R. M., Lattice dynamics of hexagonal MoS2 studied by neutron scattering. *Physical Review B* **1975,** *12* (2), 659-663.
33. Ma, Z.-X.; Dai, S.-S., Ab initio studies on the electronic structure of the complexes containing Mo—S bond using relativistic effective core potentials. *Acta Chimica Sinica* **1989,** *7* (3), 201-208.
34. Horn, H. W.; Swope, W. C.; Pitera, J. W.; Madura, J. D.; Dick, T. J.; Hura, G. L.; Head-Gordon, T., Development of an improved four-site water model for biomolecular simulations: TIP4P-Ew. *The Journal of Chemical Physics* **2004,** *120* (20), 9665-9678.
35. Abascal, J. L. F.; Sanz, E.; García Fernández, R.; Vega, C., A potential model for the study of ices and amorphous water: TIP4P/Ice. *The Journal of Chemical Physics* **2005,** *122* (23), 234511.
36. Bussi, G.; Donadio, D.; Parrinello, M., Canonical sampling through velocity rescaling. *The Journal of Chemical Physics* **2007,** *126* (1), 014101.
37. Hatch, K.; Danilowicz, C.; Coljee, V.; Prentiss, M., Demonstration that the shear force required to separate short double-stranded DNA does not increase significantly with sequence length for sequences longer than 25 base pairs. *Physical Review E* **2008,** *78* (1), 011920.
38. Boland, T.; Ratner, B. D., Direct Measurement of Hydrogen Bonding in DNA Nucleotide Bases by Atomic Force Microscopy. *Proceedings of the National Academy of Sciences of the United States of America* **1995,** *92* (12), 5297-5301.
39. Chen, J. C.; McGaughy, B. W.; Sylvester, D.; Chenming, H. In *An on-chip, attofarad interconnect charge-based capacitance measurement (CBCM) technique*, Electron Devices Meeting, 1996. IEDM '96., International, 8-11 Dec. 1996; 1996; pp 69-72.
40. Hazeghi, A.; Sulpizio, J. A.; Diankov, G.; Goldhaber-Gordon, D.; Wong, H. S. P., An integrated capacitance bridge for high-resolution, wide temperature range quantum capacitance measurements. *Review of Scientific Instruments* **2011,** *82* (5), 053904.
41. Sigalov, G.; Comer, J.; Timp, G.; Aksimentiev, A., Detection of DNA Sequences Using an Alternating Electric Field in a Nanopore Capacitor. *Nano Letters* **2008,** *8* (1), 56-63.
42. Shaeffer, D. K.; Lee, T. H., A 1.5-V, 1.5-GHz CMOS low noise amplifier. *IEEE Journal of Solid-State Circuits* **1997,** *32* (5), 745-759.
43. Cao, J.; Li, Z.; Li, Q.; Chen, L.; Zhang, M.; Wu, C.; Wang, C.; Wang, Z., A 30-dB 1–16-GHz low noise IF amplifier in 90-nm CMOS. *Journal of Semiconductors* **2013,** *34* (8), 085010.
44. Schlaak, C.; Hoffmann, P.; May, K.; Weimann, A., Desalting minimal amounts of DNA for electroporation in E. coli: a comparison of different physical methods. *Biotechnology Letters* **2005,** *27* (14), 1003-1005.
45. Feng, Y.; Zhang, Y.; Ying, C.; Wang, D.; Du, C., Nanopore-based Fourth-generation DNA Sequencing Technology. *Genomics, Proteomics & Bioinformatics* **2015,** *13* (1), 4-16.

**Supplementary information**

A MoS$_2$-based capacitive displacement sensor for DNA sequencing

A. Smolyanitsky, B. I. Yakobson, T. A. Wassenaar, E. Paulechka, K. Kroenlein

*S1. MoS$_2$ functionalization by a cytosine molecule*

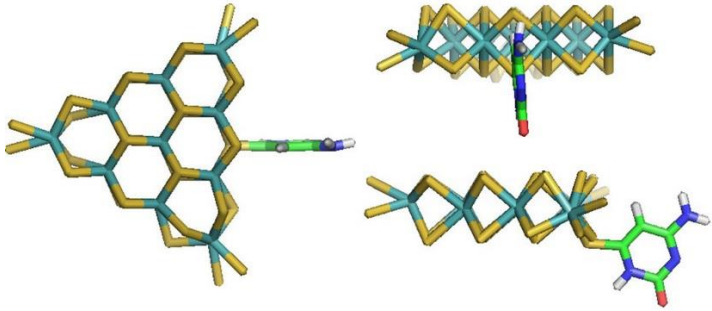

Figure S1. Views of an energy-minimized structure of a triangular MoS$_2$ cluster functionalized by a cytosine molecule.

*S2. Basic mechanical properties of simulated MoS$_2$*

The intramolecular bonded interactions for MoS$_2$ were described by the harmonic bond and inter-bond angle energy terms $E_b = \frac{k_b}{2}(r - r_0)^2$ and $E_a = k_a(\theta - \theta_0)^2$, respectively. The interaction groups and the corresponding constants are listed in Table S1.

| bond | Mo-S | $k_b = 81176.0 \ \frac{kJ}{mol \ nm^2}$ | $r_0 = 2.39$ Å |
|---|---|---|---|
| angle | Mo-S-Mo<br>S-Mo-S | $k_a = 534.16 \ \frac{kJ}{mol \ rad^2}$ | $\theta_0 = 84.3°$ |

Table S1. Bonded groups and corresponding constants describing the intramolecular interactions in simulated MoS$_2$.

As shown in Fig. S2 (a), the 2-D modulus of 126.0 N/m obtained for the MoS$_2$ model used here is close to the lower end of the range 120-180 N/m reported previously.[1-2] Note that the 0 K bending rigidity directly calculated from the 2-D modulus Y as $\gamma = Y h_m^2 / 12(1 - v^2)$ ($h_m \approx 3.12$ Å is the effective MoS$_2$ monolayer thickness and $v = 0.29$ is the Poisson's ratio)[1] yields $\gamma = 7.1$ eV, in good agreement with the finite temperature data in Fig. S2 (b).

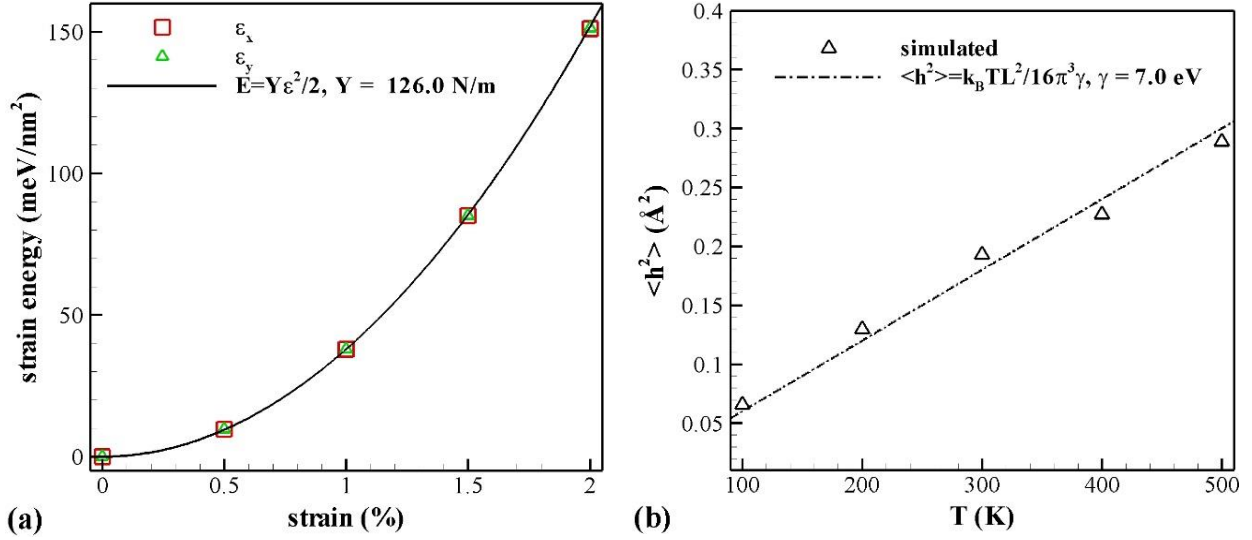

Figure S2. Simulated data and theoretical fits for the two-dimensional modulus from uniaxial stretching (a) and bending rigidity from direct simulations of ripples at a set of finite temperatures (b) of the MoS$_2$ model used in this work. A rectangular 14 nm × 15 nm MoS$_2$ sample with in-plane periodic boundaries (zigzag and armchair edges along X and Y, respectively) was used in these calculations. Theoretical fits in (a) and (b) were produced according to [3] and [4] respectively.

## S3. Capacitance perturbation due to plate deformation

A reasonable estimate is straightforward. The capacitance of a flat-plate capacitor in absence of perturbations is well known:

$$C_0 = \frac{\varepsilon \varepsilon_0 A}{d_0}, \hspace{4cm} (S1)$$

where $\varepsilon$ is the dielectric constant of the material between the plates (water, in this case), $A$ is the total plate surface area, and $d_0 \ll \sqrt{A}$ is the distance between the plates. When an out-of-plane perturbation is applied to one of the plates, the effective corresponding capacitance is subject to perturbation. Neglecting field fringing and assuming small deflections, the perturbed capacitance is

$$C = \varepsilon\varepsilon_0 \int \frac{d\Omega}{d(x,y)}, \qquad (S2)$$

where $d\Omega$ is a differential element of the flexible plate area at $(x,y)$, $d(x,y)$ is the corresponding vertical distance between the element at $(x,y)$ and the solid plate, and the integral is over the entire plate surface. We set $d(x,y) = d_0 + h(x,y)$, where $h(x,y) \ll d_0$ is the local (upwards) Z-deflection of the perturbed plate element. Hence,

$$C = \frac{\varepsilon\varepsilon_0}{d_0} \int \frac{d\Omega}{1+\frac{h(x,y)}{d_0}} \approx \frac{\varepsilon\varepsilon_0}{d_0} \int \left(1 - \frac{h(x,y)}{d_0}\right) d\Omega. \qquad (S3)$$

From Eq. (S3),

$$C \approx C_0 - \frac{\varepsilon\varepsilon_0}{d_0^2} \int h(x,y)\, d\Omega, \qquad (S4)$$

where the subtrahend is the capacitance perturbation $\Delta C$. We note that $\int h(x,y)\, d\Omega = \langle h \rangle A$, where $\langle h \rangle$ is the *average* deflection for the entire perturbed plate (in our case calculable directly from the simulated atomic positions of the membrane). Therefore, for $\langle h \rangle \ll d_0$ this estimate is reduced to finding an equivalent flat-plate capacitor with a plate separation of $d' = d_0 + \langle h \rangle$. As a result, the relative change in capacitance is independent of the deflection profile of the flexible plate $d(x,y)$, depending only on $\langle h \rangle$ and $d_0$:

$$\frac{\Delta C}{C_0} \approx -\frac{\langle h \rangle}{d_0}. \qquad (S5)$$

For a deflection $\langle h \rangle < 0$ (toward the solid plate), $\frac{\Delta C}{C_0}$ reverses sign and thus, unlike the readout scheme proposed in [5], here one can differentiate between the directions of deflection.

*S4. Ionic current perturbed by plate deformation*

The ionic leakage current between the capacitor plates is estimated similarly to the capacitance estimate presented in section S4 above. Consider KCl salt of concentration $n$, for which $\mu_K$ and $\mu_{Cl}$ are the corresponding ion mobility values. The ionic drift current via the entire capacitor is:

$$i_i = nq(\mu_K + \mu_{Cl}) \int E_z(x,y) d\Omega, \qquad (S6)$$

where $E_z(x,y) = \frac{V_0}{d(x,y)}$ is the driving field distribution throughout the plate; $V_0$ is the capacitor bias and $q$ is the ionic charge. Eq. (6) is then an analog of Eq. (2):

$$i_i = nq(\mu_K + \mu_{Cl})V_0 \int \frac{d\Omega}{d(x,y)}. \qquad (S7)$$

The approximations identical to those made in section S3 lead to the total current subject to deflection-induced perturbation:

$$i_i = \frac{nwLqV_0(\mu_K+\mu_{Cl})}{d_0}\left(1 - \frac{\langle h \rangle(t)}{d_0}\right). \qquad (S8)$$

*S5. Out-of-plane membrane pre-strain due to inter-plate electrostatic attraction*

The total attractive force between the plates carrying opposite charges of magnitude $Q$ due to bias $V_0$ is:

$$F_0 = \frac{1}{2}QV_0/d_0 = \frac{\varepsilon\varepsilon_0 wLV_0^2}{2d_0^2}. \qquad (S8)$$

For the exact dimensions of the simulated membrane of $L$ = 15.5 nm, $w$ = 4.5 nm, with $d_0$ = 1.0 nm, $\varepsilon$ = 80, and $V_0$ = 0.15 V, we obtain $F_0$ ~ 555 pN, *distributed throughout the entire membrane area*. Maximum deflection of the membrane with fixed edges and an out-of-plane load $F_0$ homogeneously distributed throughout the membrane is $h_{max} = \frac{F_0 L^3}{384 \gamma w}$, where $\gamma$ is the bending rigidity of the material – half the deflection for the case of $F_0$ *concentrated at the center*. Given that the deflection in response to ~75 pN (concentrated at the center) arising from hydrogen bonds has already been simulated at ~0.5 Å, and finally noting that $\langle h \rangle \approx 0.5\, h_{max}$ (see Fig. 2 (a)), the pre-deflection due to inter-plate attraction is $\langle h \rangle_0 \approx 0.5$ Å $\times$ 0.5 $\times$ (555/75) = 1.85 Å, merely suggesting that the geometrically selected plate separation (as calculated at the supported ends of the membrane) should be $d_0 + \langle h \rangle_0$ = 1.185 nm to yield $d_0$ = 1.0 nm used in our electrostatics estimates. *It is critical to realize that the small plate separation and the bias voltage of 150 mV were selected only to produce a reasonable capacitance value and the corresponding charge perturbation $\delta Q = V_0 C_0 \langle h \rangle / d_0$ of at least a few electron charges, given the small system size and thus the quantized nature of the charge transfer process.*

By rescaling the device toward more realistic dimensions, this pre-deformation effect can be greatly diminished. For example, with $L$ = 100 nm, $w$ = 75 nm, $d_0$ = 25 nm, the new capacitance value is estimated at 212 aF, while the voltage can be selected at 50 mV, which leads to an inter-plate force of only 10 pN. The expected average deflection $\langle h \rangle$ from breaking a C-G pair for this membrane is estimated at ~ 2 nm $\ll d_0$, thus remaining within the approximations made above and in the main text.

*S6. Remarks on displacement current and ionic leakage*

The data in Fig. 3 and analytical calculations can be used to provide a rough estimate of the threshold electrolyte concentration, beyond which one cannot rely on the capacitive effect in a DC circuit. Per each deflection event, two displacement current peaks occur, according to $i_d(t) = V_0 \frac{dC(t)}{dt}$: one during while the membrane deflects, and one when it snaps back. The timescale of the former is associated with the translocation rate, while the timescale of the latter process is governed entirely by the membrane material properties and size, as well as the damping properties of the solvent. For both processes, the displacement current is proportional to *the rate, at which the capacitance changes* and in the following we only consider the deflecting part, while the snap-off process, leading to the second peak, can be evaluated in a similar manner. According to Eq. (S5), the current peak magnitude (in absence of noise) is:

$$i_{d,peak} = \frac{V_0 C_0}{d_0} \frac{d\langle h \rangle}{dt} = \frac{\varepsilon \varepsilon_0 w L V_0}{d_0^2} \frac{d\langle h \rangle}{dt}. \qquad (S9)$$

By noting that $\langle h \rangle \approx h_{max}/2$ (see Fig. 2 (a)) and $\frac{dh_{max}}{dt} \approx u_t$, where $u_t$ is the translocation speed, we obtain:

$$i_{d,peak} \approx \frac{\varepsilon \varepsilon_0 w L V_0 u_t}{2 d_0^2}. \qquad (S10)$$

The ionic current mostly contributes a considerable baseline shift, but also fluctuates as a result of deflection (according to Eq. (S8)) – proportional to $\langle h \rangle$ and not $d\langle h \rangle/dt$. The magnitude of this current fluctuation is of interest. Its peak occurs at the maximum average deflection $\langle h \rangle_{max}$ (immediately prior to base-pair breakage) and the corresponding magnitude is:

$$i_{i,peak} \approx \frac{n w L q V_0 (\mu_K + \mu_{Cl}) \langle h \rangle_{max}}{d_0^2}. \qquad (S11)$$

The peaks given by Eqs. (S10) and (S11) differ by phase, but their magnitudes can be directly compared to determine the electrolyte concentration, at which the displacement current peak from deflecting the membrane no longer dominates. By setting $\frac{i_{i,peak}}{i_{d,peak}} > 1$, we obtain:

$$n > \frac{2\varepsilon\varepsilon_0 u_t}{q(\mu_K + \mu_{Cl})\langle h \rangle_{max}}, \quad (S12)$$

where membrane size dependence enters only via $\langle h \rangle_{max} \propto L^3/w$. *Given direct proportionality to $u_t$, capacitance-based detection favors fast translocation in the DC-bias case.* Assuming ionic mobility of the single-charged K$^+$ and Cl$^-$ ions of ~ $8 \times 10^{-8} \frac{m^2}{V \cdot s}$, $\langle h \rangle_{max} = 0.5$ Å, and $u_t = 5\ cm/s$, we estimate that with $n > 0.1$ M, ionic leakage starts to dominate charge transfer in the system, masking *the deflection-induced current peak (the discussion above is not applicable to the peak due to membrane snapping off, because it is not affected by the translocation rate)*. With increasing membrane dimensions (and thus generally increasing $\langle h \rangle_{max}$) this threshold decreases as $1/\langle h \rangle_{max}$, which may be a consideration for the electrolyte concentration management effort.

# References

1.  Jin-Wu, J.; Zenan, Q.; Harold, S. P.; Timon, R., Elastic bending modulus of single-layer molybdenum disulfide (MoS 2 ): finite thickness effect. *Nanotechnology* **2013,** *24* (43), 435705.
2.  Jiang, J.-W.; Park, H. S.; Rabczuk, T., Molecular dynamics simulations of single-layer molybdenum disulphide (MoS2): Stillinger-Weber parametrization, mechanical properties, and thermal conductivity. *Journal of Applied Physics* **2013,** *114* (6), 064307.
3.  Singh, S. K.; Neek-Amal, M.; Costamagna, S.; Peeters, F. M., Rippling, buckling, and melting of single- and multilayer MoS2. *Physical Review B* **2015,** *91* (1), 014101.
4.  Gao, W.; Huang, R., Thermomechanics of monolayer graphene: Rippling, thermal expansion and elasticity. *Journal of the Mechanics and Physics of Solids* **2014,** *66*, 42-58.
5.  Paulechka, E.; Wassenaar, T. A.; Kroenlein, K.; Kazakov, A.; Smolyanitsky, A., Nucleobase-functionalized graphene nanoribbons for accurate high-speed DNA sequencing. *Nanoscale* **2016,** *8* (4), 1861-1867.